\documentclass[linenumbers]{emulateapj}
\usepackage{graphicx} 
\usepackage{lipsum}
\usepackage{graphicx}
\usepackage{hyperref}

\usepackage{amsmath}
\usepackage{longtable}

\shorttitle{Time Dependent Photoionization}
\shortauthors{Sadaula et al.}

\graphicspath{{./}{figures/}}
\begin{document}
\title{Time Dependent Photoionization Modeling of Warm Absorbers: High-Resolution Spectra and Response to Flaring Light Curves}

\author{Dev R Sadaula}
% \affiliation{Western Michigan University \\
% 1903 W Michigan Ave,\\
% Kalamazoo, MI 49008, USA}
\affiliation{NASA Goddard Space Flight Center \\
8800 Greenbelt Rd\\
Greenbelt, MD 20771, USA}
\affiliation{University of Maryland Baltimore County\\
1000 Hilltop Circle\\
Baltimore, MD 21250, USA}
% 1903 W Michigan Ave,\\
% Kalamazoo, MI 49008, USA}

% \author{Manuel A Bautista}
% \affiliation{Western Michigan University \\
% 1903 W Michigan Ave,\\
% Kalamazoo, MI 49008, USA}

% \author{Javier A Garc{\'\i}a}
% \affiliation{California Institute of Technology \\
% 1200 E California Blvd,\\
%  Pasadena, CA 91125, USA}
% \affiliation{ Dr. Karl Remeis-Observatory and Erlangen Centre for Astroparticle Physics\\
% Sternwartstr. 7, \\
% D-96049 Bamberg, Germany}

\author{Timothy R Kallman}
\affiliation{NASA Goddard Space Flight Center \\
8800 Greenbelt Rd\\
Greenbelt, MD 20771, USA}

\begin{abstract}
Time dependent photoionization modeling of warm absorber outflows in active galactic nuclei can play an important role in understanding the interaction between warm absorbers and the central black hole. The warm absorber may be out of the equilibrium state because of the variable nature of the central continuum. In this paper, with the help of time dependent photoionization modeling, we study how the warm absorber gas changes with time and how it reacts to changing radiation fields.  Incorporating a flaring incident light curve, we investigate the behavior of warm absorbers using a photoionization code that simultaneously and consistently solves the time dependent equations of level population, heating and cooling, and radiative transfer. We simulate the physical processes in the gas clouds, such as ionization, recombination, heating, cooling, and the transfer of ionizing radiation through the cloud. We show that time dependent radiative transfer is important and that calculations which omit this effect quantitatively and systematically underestimate the absorption.  Such models provide crucial insights into the characteristics of warm absorbers and can constrain their density and spatial distribution.
\end{abstract}


\section{Introduction} \label{sec:intro}
Active galactic nuclei (AGNs) are characterized by energetic phenomena, including the emission of radiation across the electromagnetic spectrum, ranging from radio waves to X-rays.  This, together with the existence of a supermassive black hole at the center, hints at the existence of gas flows, both inward and outward, near the center.  Such gas flows can be studied observationally by observing the absorption and emission produced by this gas. Such absorption includes warm absorbers (WAs), which intercept and absorb specific wavelengths of light, resulting in distinctive absorption features in the observed X-ray spectra. The study of these absorption features provides valuable insights into the physical properties and dynamics of the absorbing gas.

Warm absorbers experience temporal fluctuations in the radiation continuum coming from the central region of AGN. The brightness of the AGN varies over time due to flare events such as alterations in the black hole's accretion rate or interactions with its surroundings. A key component of the study of warm absorbers is comprehending how they react to these flares.

The effect of continuum variability is most important when the time scales of the intrinsic microscopic processes characterizing the ionization state and temperature are comparable with the variability time scale of the central source. If so, it is important to perform time dependent calculations. AGN variability has been extensively observed and reveals significant variability ($\geq 20 \% $) on timescales as short as $\sim 10^3$ s \citep{sil16}.  Time dependent photoionization has been explored previously in various astrophysical contexts. Examples include  the time dependent cooling in photoionized intergalactic gas \citep{gna17}; time dependent radiative transfer equation as applied to  AGN outflows and gamma-ray bursts \citep{mat12}; time evolution of H II regions, including the time dependence of radiation transfer \citep{gar13}; and recent models for time dependent photoionization modeling of AGN outflows by us \citep{sadaula23} and by \cite{rog22,luminari22}

Time dependent photoionization analysis of a gas exposed to a variable source can be used as a density diagnostic.  When the illuminating flux varies, the mean ionization state of a gas responds on a timescale dependent on the flux and the gas density. For a gas initially in equilibrium, the flux and density are proportional to each other via the ionization parameter. Gas with low density reacts slowly to variations in the ionizing flux; response to a varying flux will be averaged over a long timescale. When the flux changes for a gas with a high density (e.g., $\geq$  $10^{10}$ cm$^{-3}$ for the cases of interest here), the gas reacts swiftly and largely maintains the equilibrium state. Gas of intermediate gas density reacts to the flux change but with a delay or smearing in time. Based on this delay, it is possible to deduce the density of the gas \citep{nic99}.

In our previous work \citep{sadaula23} (Hereafter, we refer to this as paper 1), we modeled time dependent ionization of warm absorbers in the idealized situation in which the gas is exposed to a flux that varies with a simple step-up from a lower to a higher flux, or a step down from a higher to a lower flux.  We also used a relatively crude energy grid, and so produced spectra with a resolution appropriate to low-resolution detectors, i.e., resolving power $E/ \Delta E\sim 10$.  In the present study, we explore the behavior of warm absorber response to gas exposed to flaring light curves. We present the details of the time evolution of the relevant quantities: ion fractions, temperature, and the transmitted spectrum. The transmitted spectra are calculated using a full XSTAR \citep{kall01} atomic database. The ion fractions obtained from time dependent photoionization calculations are used to simulate the high-resolution transmission spectra and therefore produce spectra with resolving power $E/\Delta E\sim 1000$. In addition, we examine the effects of modeling the outflow considering the absorber as if it had a single spatial zone. It implies the exclusion of the radiative transfer equation from the set of time dependent equations. Even though this method is computationally efficient compared to the multi-zone approach and gives nearly the same results for relatively smaller columns, it departs noticeably for a higher column of absorbers. We demonstrate the difference between single- and multi-zone modeling.   

In the rest of the paper, we present the model and assumptions in Section 2, the results for the step-up model in Section 3, the result for the flare model with multiple spatial zones in Section 4, and the result for the flare model for a single zone cloud in Section 5. Then, we present our conclusions. The essential equations, assumptions, and numerical methods are presented in the Appendix since much of this is a summary of our previous work \citep{sadaula23}.

\section{Models and Assumptions}
In paper 1, we studied a range of parameter values affecting warm absorber models:   gas densities, ionization parameters, column densities, and continuum variability light curves. Even though many types of models were presented in paper 1, we limited our calculation to relatively low spectral resolution. In addition, all the models were run for a light curve consisting of a single sudden change in the source luminosity, either upward or downward. We term these step-up or step-down models. The results from paper 1 give insight into the evolution of the gas in response to a sudden change in the ionizing luminosity and the importance of the time dependent calculation.  However, in order to compare our model calculations with observations, we need to produce a high spectral resolution and consider more realistic light curves for the illuminating flux.

\subsection{Ionization Parameter}
The ionization parameter is a convenient description of the ionization and temperature of a photoionized gas under the assumption of equilibrium. If so,  the ionization parameter can be defined as,
\begin{equation}{\label{xi}}
    \xi = \frac{L_{ion}}{n_HR^2} = \frac{4\pi F_{ion}}{n_H}    
\end{equation}
where $L_{ion}$ is the ionizing luminosity of the source, $F_{ion}$ is the ionizing flux, and both of these quantities are integrated from 1-1000~Ry. $R$ is the distance of gas from the ionizing source. This definition was first introduced by \cite{tarter69}. The other way of defining ionization parameter is $U=\int_{\varepsilon_{1}}^{\varepsilon_2} L_{\varepsilon}  / (\varepsilon 4 \pi R^2 n_H c) d\varepsilon,$ where, $L_{\varepsilon}$ is the ionizing luminosity per energy interval and $c$ is the speed of light. $U$ and $\xi$ are easily convertible to each other for a given spectral energy distribution of the ionizing continuum.  For the spectrum shape we adopt here, i.e., a single power law from 1 - 1000 Ry with energy index -1, the conversion is $U=\xi/56.64$ erg cm s$^{-1}$. The higher the ionization parameter, the more ionized the gas and vice-versa. Warm absorber components generally exist either in the low ionization state of $ \xi\sim 1$ erg cm s$^{-1}$ or high ionization state of $\xi \sim 100$ erg cm s$^{-1}$ \citep{laha14}.

It is important to clarify the usage of luminosity and flux. These two quantities are interconnected through equation \ref{xi}. Luminosity refers to the total ionizing luminosity emitted by the source, while flux represents the luminosity of the source divided by the area of the sphere at the specific point of interest within the cloud. The flux is a convenient description of the radiation strength and its effects on gas locally, while luminosity is a useful measure of the global energetics of the central source.

\subsection{Models}
The free parameters in our models include hydrogen number density ($n_H$; hereafter hydrogen gas density unless otherwise stated), hydrogen column density ($N_H$; hereafter column density means hydrogen column density otherwise stated), ionization parameter ($\xi $), the shape of the incident radiation as described by the spectral energy distribution (SED), and element abundances. We have created models for warm absorber outflows depending mainly on the type of incident light curve. These are 1) a step-up model, 2) a flare model with multi-zone consideration, and 3) a flare model with single-zone consideration. We refer to the step-up model as model 1, a flare model with multi-zone consideration as model 2, and a flare model with single-zone consideration as model 3 in the subsequent sections for simplicity. The details of these models are described in the corresponding sections.

\subsection{Assumptions}
Since this work is an early step toward exploring photoionized plasma in non-equilibrium conditions, we attempt to make our results general and illustrate the general behavior of time dependence.  Therefore we have made several key approximations, which are as follows:
\begin{enumerate}
    \item We approximate the gas to be static, with no expansion of the gas and no velocity gradient. For a supersonic flow, such as in a warm absorber, the bulk forces, whatever they are, dominate over internal dynamics in the cloud \citep{proga22}.  If so, our models can be applied by adding a blueshift.  We acknowledge this effect can be important, but it is not included in the results presented here. We can calculate models which can be refined for specific situations, and we plan to do so in future work.
    \item We assume the outflowing gas has constant density throughout. This assumption is partially justified by the physical thickness of the cloud, which is less than $10\%$ with respect to the distance from the ionizing source, as shown in paper 1. If so, the imprint of any large-scale spherical flow will be small. However, we acknowledge that there may be a density gradient in the case of warm absorbers associated with dynamical effects such as ram pressure, which will be considered in future work.
    \item We have not included emission in our radiative transfer calculation. This approximation is justified if the covering fraction of the absorber relative to the central source is small; this appears to be the case for most observed warm absorbers. For most observed warm absorber spectra, emission features are weak or absent  \citep{kas01,kas02}. However, including emissions is on our list of tasks for future work since the emission is likely to be important in other time dependent applications.
\end{enumerate}

\subsection{High Resolution Spectra}
Time dependent calculations require the simultaneous solution of the equations describing the ionic level populations, temperature, and radiation field for all level populations,  spatial zones, and photon energies.  This corresponds to a very large number of simultaneous ordinary differential equations; the standard {\sc xstar} database has $\sim 3 \times 10^4$ energy levels. Running these full schemes is extraordinarily time-consuming and sometimes gives unstable solutions if the number of equations is too large. This necessitates a simplification of this system of equations to make it tractable.  That is, we simplify the energy level structure of the ions in the gas. To do this, we have created an atomic database that describes each ion using three levels: the ground level, one excited bound level, and the continuum (ionized) level.  For the one bound excited level, we adopt an {\it ad hoc} description:  the level energy is 0.8$\times E_{th}$ where $E_{th}$ is the ionization potential of the ion. We include electron impact collisional excitation to the level for the ions H$^0$, He$^0$, and He$^+$ with an effective collision strength (upsilon) with a value of 1.4 to 2.7 depending upon temperature. The excited level is a pseudo level, and the energy of that level is chosen in such a way that the temperature vs. ionization parameter ($\xi$) curve can agree as closely as possible with the temperature vs. ionization parameter ($\xi$) curve obtained from the full {\sc xstar} databases.  This is illustrated in Figure 2 of paper 1 \citep{sadaula23}. We include all of the ions that are in {\sc xstar} in the newly created database, though the simulations in this paper include only a subset of them. Paper 1 \citep{sadaula23} gives more details about this small database.

In addition, we are limited by the number of time, position, and energy grid points that can be included in our computation while solving the discretized time dependent equation of level population, energy balance, and radiative transfer. The energy grid we can use for such calculations is inadequate for calculating high-resolution spectra. To address this, we created a two-step technique.

\begin{enumerate}
    \item We first run the time-dependent photoionization code with reasonable energy, position, and energy grid numbers using the small subset of the XSTAR database, with each ion represented by two bound levels plus a continuum.
    \item The ground-level populations, which are mainly responsible for the absorption of the ionizing radiation of the source, are then used to produce high-resolution spectra using the spectral fitting model package  WARMABS~\footnote{https://heasarc.gsfc.nasa.gov/docs/software/xstar/xstar.html}. WARMABS uses the full XSTAR database and can handle arbitrary energy grid spacing. We then use the opacities from WARMABS to calculate the emergent absorption spectrum via the formal solution to the radiative transfer equation (including the time delays associated with light travel time).  In this way, we can produce high-resolution spectra.
\end{enumerate}

\section{Results: Model 1, step-up model} \label{sec4}
Paper 1 \citep{sadaula23} comprehensively discussed the general behavior of time dependent photoionized models. It covered various aspects such as ion fractions, temperature, and transmitted spectra. The current study expands upon our previous work by incorporating a high-resolution spectrum. We have done this using the full XSTAR database and adding a more dense energy grid.  We also track the evolution of the gas until it returns to a state of high flux equilibrium. Model 1 illustrates the progressive changes observed in the high-resolution spectra when exposed to a step-up flare incident light curve.

As was true for our models in paper 1, we capture the time dependence of the propagation of the radiation through the cloud by solving the time dependent radiation transfer equation.  We employ a multiple-zone approach by dividing the gas cloud into distinct slices and solving the radiative transfer equation (\ref{transfer} presented in the appendix) on this grid using a finite-difference representation.

\subsection{Model Parameters}
Model 1 has gas density $10^7$ cm$^{-3}$; column density is $9.0 \times 10^{22}$ cm$^{-2}$, initial ionization parameter ($\xi_1$) is $\sim 50$ erg cm s$^{-1}$, incident ionizing spectral shape of a power law with energy index -1, initial luminosity of the source $10^{44}$ erg~s$^{-1}$ and includes elements: H, He, C, O, Si, and Fe. The incident ionizing radiation flux suddenly increases by a factor of three at the beginning of the simulation.

\subsection{Step-up light curve}
\begin{figure*}[ht]
\centering
\includegraphics[width=0.8\textwidth]{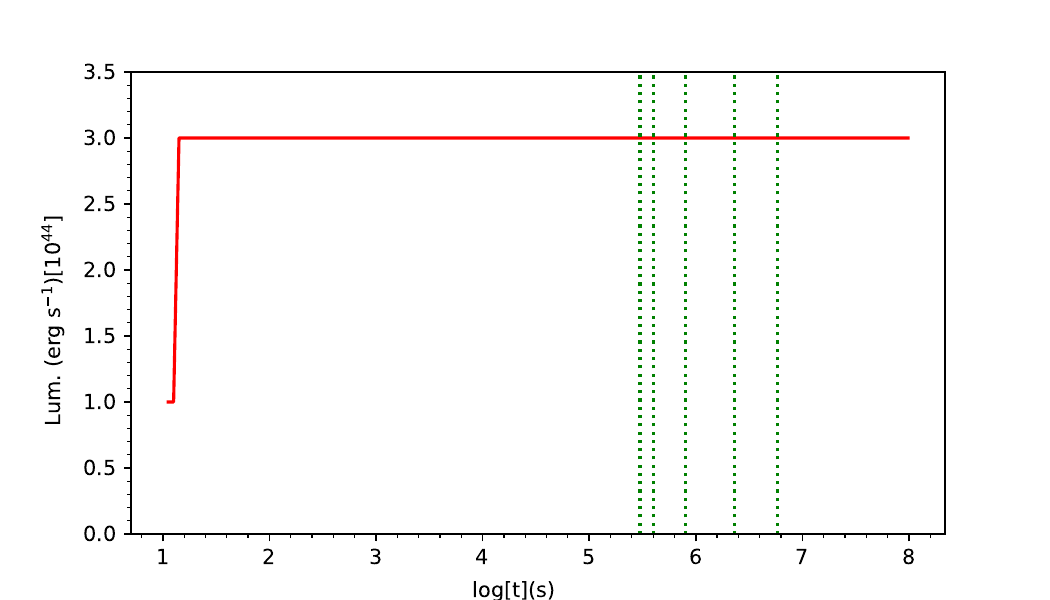}
\caption{The light curve at the face of the cloud(red line) in units corresponding to source luminosity. The source luminosity was changed by a factor of three to a higher value ($3.0 \times 10^{44}$ erg s$^{-1}$). The five vertical green dotted lines are times at a high flux state. We have chosen them to illustrate the evolution of the model absorption spectrum.}
\label{lc_step}
\end{figure*}

In Figure \ref{lc_step}, we present the input light curve for model 1. The initial luminosity of the source is set at $10^{44}$ erg s$^{-1}$, and it is subsequently increased to $3.0 \times 10^{44}$ erg s$^{-1}$ at $\sim$ 20 s. We have included five vertically dotted green lines to highlight specific times for which we will display the spectra. These times are chosen such that the changed radiation field has completely emerged through the cloud. The simulation for this model is conducted for $10^8$ s.

\subsection{Absorption Spectra}
\begin{figure*}[ht]
\centering

\includegraphics[height=22cm, width=16cm]
% {step_spec_col_23_final_blue_fac_3_2line_ratio.pdf}
{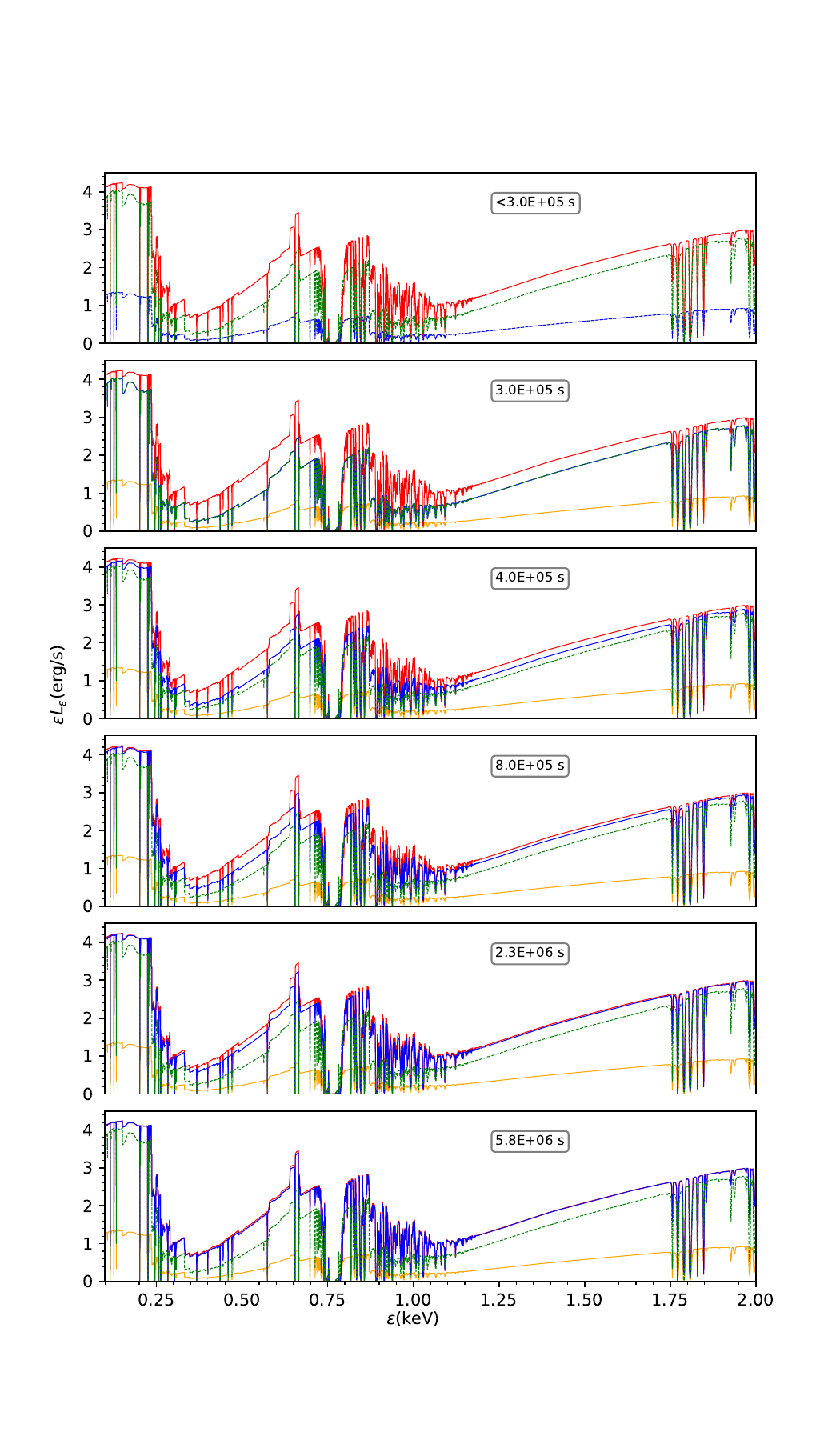}
\caption{Figure showing snapshot of the evolution of absorption spectra for a step-up model with gas density $n=10^7$ cm$^{-3}$. The y-axes represent the transmitted luminosities ($\varepsilon L_\varepsilon$), and the x-axes represent the photon energies ($\varepsilon$). The units of $\varepsilon L_\varepsilon$ are $10^{43}$ erg~s$^{-1}$. The orange curves represent lower flux equilibrium spectra, the red curves show the high flux equilibrium spectra and the blue curves show the evolution of transmitted spectra during the simulation. We have also plotted the green spectra for reference, obtained by multiplying a low flux equilibrium spectrum by a factor of three.  The times in the box correspond to the vertical dashed green lines on the light curve in figure \ref{lc_step}.}
\label{spec_step}
\end{figure*}

Figure \ref{spec_step} displays the evolution of transmitted spectra for model 1.  Each panel illustrates the transmitted spectrum at a specific moment in time. It is important to note that the time it takes for the radiation to propagate from the source to the illuminated face of the absorber is not considered. The simulation begins with the arrival of the changed flux at the face of the cloud. The initial spectrum showcases the absorption features and characteristics before the increase in luminosity occurs. Over time, as the higher flux propagates through the absorber, the transmitted spectra in subsequent panels exhibit changes in the absorption features. These changes are a consequence of the altered ionization and excitation within the gas due to the increased luminosity of the ionizing source. The simulation is primarily on the effects of the sudden change in luminosity once it reaches the surface of the cloud, the illuminated face. Figure~\ref{spec_step} visually represents the evolving transmitted spectra and their response to the abrupt increase in ionizing source luminosity. It provides information about the time dependent behavior of the absorber and the impact of changes in the incident flux on its observed spectral features.

The spectra presented in figure \ref{spec_step} constitute many absorption features. The absorption features spanning the energy range $\simeq$ 0.2 -- 0.7 keV primarily arise from the absorption of carbon (C V and C VI), silicon, and iron L shell ions.  The second group of absorption features,  located $\simeq$ 0.75 keV, predominantly results from iron ions.  The absorption edge at $\simeq$ 1 keV, is primarily influenced by iron and oxygen. The interaction between these ions and incident radiation gives rise to the distinctive absorption edge observed in this energy range. Throughout the spectrum, absorption lines arise from a wide range of ions associated with various elements. 
%These lines provide further insights into the ionization states, composition, and physical properties of the gas cloud under investigation. 
In each panel, three different spectra are presented: the red curve represents the high flux equilibrium spectrum, the orange spectrum corresponds to the low flux equilibrium spectrum, and the blue curve represents the time dependent spectrum.  The green curves are obtained by multiplying the low flux equilibrium spectrum by three. These reference spectra help to visualize how the ionized absorber changes its opacity over time due to the sudden change in the luminosity of the source from low to high. These spectra allow for a comparative analysis of the evolving absorption features and their variations in response to changing flux conditions. By examining the figure and the different spectra presented, it is possible to gain a deeper understanding of the time dependent behavior of the absorption features, as well as the contributions of different ions and elements to the overall spectrum.

The first panel of figure \ref{spec_step} corresponds to the case before the changed radiation field arrives at the back of the cloud. The time it takes light to cross this cloud for this model is $\simeq 3\times10^5$ s. This is why the orange and blue spectra overlap until this time. When the altered radiation arrives at the back of the cloud, the blue spectrum changes towards the red spectrum and overlaps with the green spectrum because the gas has not yet had a chance to change its ionization state, as seen in the second panel. Since the opacity of the gas cloud decreases over time due to a high ionizing radiation field, the gas becomes more transparent, and the spectrum starts to move closer to the red curve and farther from the green curves. This model takes $\sim 5.8\times 10^6$ s for the gas to reach an equilibrium state corresponding to the high flux. This is the time including the effect of photoionization and thermal equilibrium. The timescale of thermal equilibrium is much longer than the photoionization equilibrium \citep{sadaula23}. %This equilibrium time scale  depends upon the density of the gas \citep{nic99}. 

% We have produced a ratio plot shown in figure \ref{ratio_step} to show how the opacity of the absorbers changes over time. For this, we plotted the two different ratios vs. energy at different times. The blue curves are the ratio of the spectra that come from time dependent photoionization equilibrium calculation (blue spectra in figure \ref{spec_step}) to the high flux equilibrium spectra (red spectra in figure \ref{spec_step}) while the green curves are the ratio of three times the low flux equilibrium spectrum to the final high flux equilibrium spectra.

\begin{figure*}
\includegraphics[height=22cm, width=16cm]
{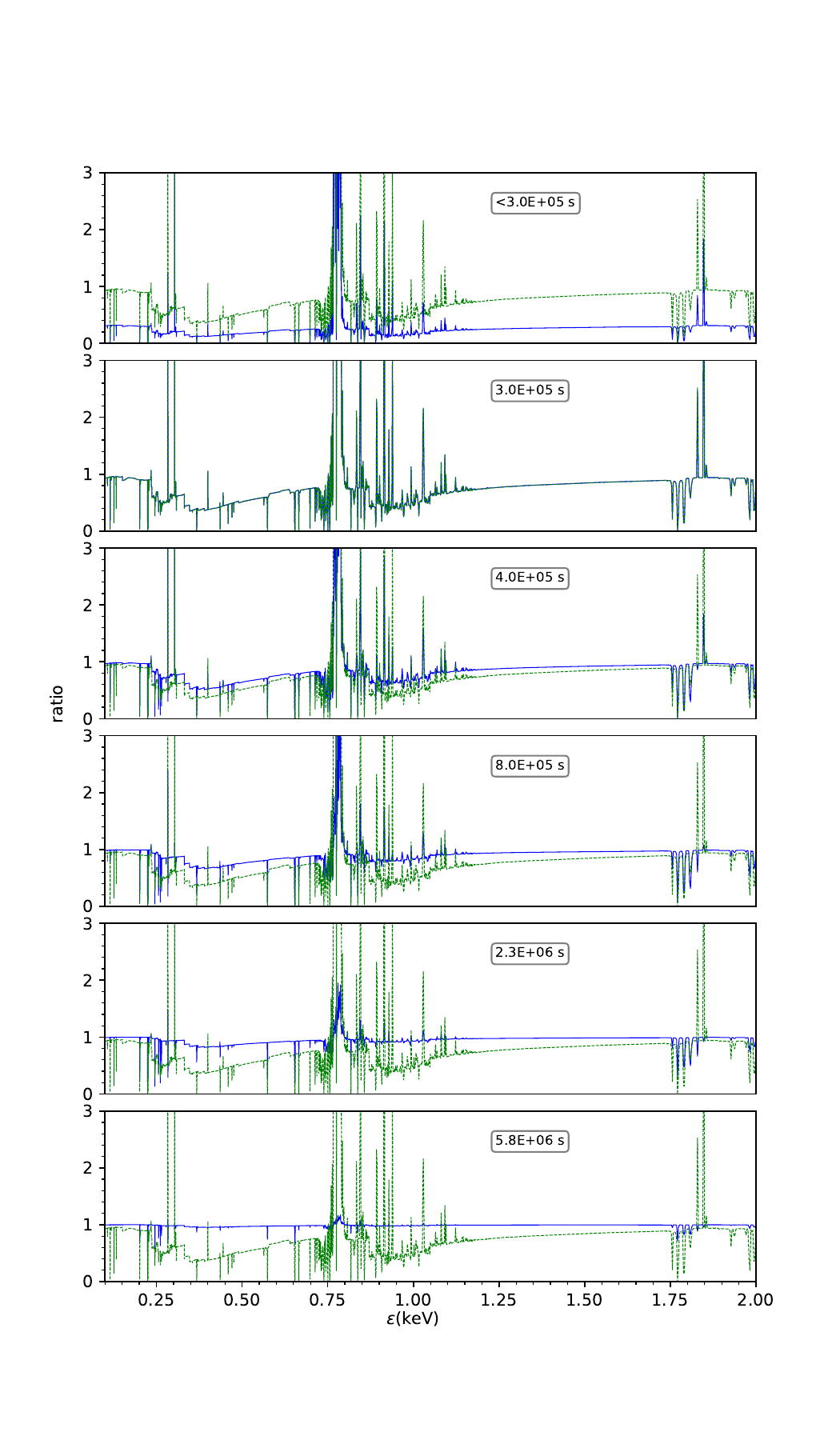}
% {step_spec_col_23_final_blue_fac_3.pdf}
\caption{Figure showing snapshot of the absorption spectra evolution by plotting the spectral ratio for model 1. The blue line is the ratio of time dependent and the final high flux equilibrium spectra, while the green is the ratio of three times the low flux equilibrium spectra to final high flux equilibrium spectra (this is plotted for reference). The times in the box correspond to the vertical dashed green lines marked on the light curve in figure \ref{lc_step}. The details are given in the description.}
\label{ratio_step}
\end{figure*}

We have generated a ratio plot, depicted in Figure \ref{ratio_step}, to illustrate the dynamic evolution of absorber opacity. To achieve this, we created two distinct ratios against energy at various time points. The blue curves represent the ratios of spectra resulting from time-dependent photoionization equilibrium calculations (blue spectra in Figure \ref{spec_step}) to those of the high-flux equilibrium spectra (red spectra in Figure \ref{spec_step}). On the other hand, the green curves denote the ratios of three times the low-flux equilibrium spectrum (green spectra in Figure \ref{spec_step}) to the ultimate high-flux equilibrium spectra.

The green line within Figure \ref{ratio_step} serves as a reference line for comparing alterations in transmitted absorption spectra at various times during the simulation. The blue curve in the first panel corresponds to the time before the altered radiation field reached the back of the cloud. This is why the blue and green curves differ in absorption features and continuum. In the second panel, both of these curves overlap each other. The altered radiation field has just arrived at the back of the cloud by this time. In the rest of the panel, the green curve starts to depart at absorption features because the gas starts to be ionized and becomes transparent as time passes. The opacity of the gas changes until the blue line becomes nearly horizontal, and the ratio becomes one. This is the final flux ionization state of the gas.

\section{Results: Model 2, flare model with multiple zones}
In this section, we investigate the response of the gas cloud to a flare-like event, aiming to simulate scenarios observed in certain variable AGN where a sudden increase in luminosity occurs, followed by a return to the initial state. This behavior may be attributed to a rise in accretion rate onto the central black hole. As was the case with model 1, to capture this phenomenon, we employ a multiple-zone approach by dividing the gas cloud into distinct slices and incorporating the radiative transfer equation into our model.

By considering the radiative transfer equation, we account for the interaction of radiation with the gas in each zone. This allows us to examine how the radiation propagates through the cloud and affects its physical and ionization properties. The multiple-zone approach provides a more comprehensive and detailed description of the gas cloud's response to the flare, enabling us to capture the spatial variations and dynamic processes occurring within the cloud. This enables us to study the evolution of absorption features, emission lines, and other spectral characteristics as the flare propagates through the cloud.

The inclusion of radiative transfer in our model allows us to account for the time delays associated with the propagation of radiation through the cloud. As the flare radiation travels through the different zones, it interacts with the gas, leading to ionization changes and the emergence of absorption and emission features at different stages of the flare evolution.

By analyzing the time dependent spectra obtained from the multiple zone approach, we can explore the evolution of the absorption features and the corresponding ionization states of various elements in the cloud. This enables us to investigate the interplay between the radiation field, ionization processes, and the physical conditions within the cloud during the flare event.

In the following sections, we present the detailed results of our multiple-zone modeling approach, including the time dependent evolution of the spectra, the variation of absorption features, and the ionization states of different elements across the cloud. These results provide insight into the behavior of warm absorbers in response to flaring events, shedding light on the underlying physical processes in AGN and their impact on the observed spectra.

\subsection{Model Parameters}
Model 2 has gas density is $10^7$ cm$^{-3}$, initial luminosity of the source $10^{44}$ erg s$^{-1}$, initial ionization parameter ($\xi_1$) is $\sim 81$ erg cm s$^{-1}$, column density is $10^{22}$ cm$^{-2}$, the incident ionizing spectral shape of power law with energy index -1, and include elements: H, He, C, N, O, Ne, Mg, Si, S and Fe. The incident ionizing radiation flux suddenly increases by a factor of ten.  The cloud in this model is divided into 60 spatial zones. The simulation is run for $10^5$ s. 

\subsection{Flare Light Curves}
\begin{figure*}[ht]
\centering

\includegraphics[width=0.8\textwidth]{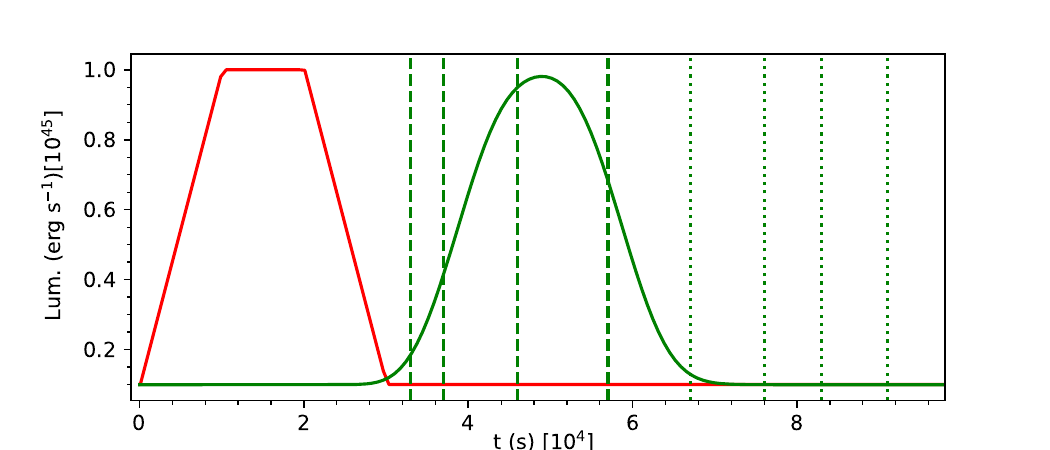}
\caption{The light curves at the face (red curve) and at the back of the cloud (green curve) for model 2. The source luminosity changed by a factor of ten. The variability time of the flare is $3.0 \times 10^4$ s. The first four vertical green dashed lines are at different times during the flare, and the last four green dotted lines are at a time when the luminosity settles at a low state.}
\label{lc_flare}
\end{figure*}

Our study considers the incident light curve, as depicted in Figure \ref{lc_flare}. The red curve represents the incident light curve at the illuminated face of the cloud, whereas the green curve is at the back of the cloud.  This incident light curve exhibits three distinct regions, each with its own characteristics. The first region corresponds to a linear increase in source luminosity until $10^4$ s. During this phase, the flare is in its initial rising state, gradually increasing in intensity. The second region of the light curve represents a high luminosity state that remains constant for an additional $10^4$ s. This phase corresponds to the peak of the flare, where the source maintains a steady high luminosity. After the high state, the third region shows a linear decrease in flux over a duration of $10^4$ s. This marks the declining phase of the flare, where the luminosity gradually decreases with time until it reaches a low luminosity state. Overall, the entire variability time of the flare is $3\times10^4$ s, which encompasses the rising, peak, and declining phases.

In our multiple-zone modeling approach, we consider the propagation of the flare through the gas cloud. As the flare progresses deeper into the cloud, its temporal behavior changes, resulting in a smoother variation compared to the initial light curve. This effect can be observed by examining the green line in Figure \ref{lc_flare}, which represents the light curve obtained at the back of the cloud.

The time for the flare to travel through the cloud is determined by the sum of three components: the variability time of $3\times10^4$ s, the light travel time, and the propagation time. The light travel time is the time it takes for the flare to travel from the front of the cloud to the back in the absence of any absorption. In our model, the light travel time is estimated to be $\sim 3.3\times10^4$ s. Therefore, the total time for the flare to traverse through the entire cloud is $\sim 7.0\times10^4$ s, as shown in figure \ref{lc_flare}. The propagation time primarily comes from the variable absorption process of ionizing radiation within the cloud. It depends on the ratio between the number of atomic species undergoing ionization within a specific spherical shell and the rate of ionizing photons emitted by the source in case of increasing source luminosity. The propagation time, therefore, depends upon the ionization and recombination time of the ions. For a more comprehensive and quantitative description of the propagation time, please refer to paper 1 \citep{sadaula23}. For reference, the photoionization and recombination times of some important ions are presented in the table \ref{time_tab}. These times are calculated at the illuminated face of the cloud, using equilibrium calculation for the low luminosity value of the source. These times are taken from the simulation and hence include some errors. Most of the error appears due to the coarse energy grids and the simplified database we have used.

\begin{table*}[t]
    \centering
    % \begin{tabular}{|l|l|l|l|l|l|l|l|l|l|l|}
    \begin{tabular}{|c|c|c|c|c|c|c|c|}
    \hline
    Ions & C VI & O VII& O VIII& Si XIII& Si XIV& Fe XV& Fe XVI\\\hline
    Photoionization time (s) &$1.7\times 10^3$&$3.1\times 10^3$&$1.0\times 10^0$&$1.2\times 10^2$&$3.7\times 10^2$&$2.7 \times 10^1$&$2.6 \times 10^1$\\\hline
    Recombination time (s) &$2.0\times 10^4$&$1.4\times 10^4$&$2.1\times 10^4$&$5.0\times 10^3$&$2.0\times 10^3$&$3.3\times 10^2$&$2.2\times 10^2$\\\hline
    
    \end{tabular}
    \caption{Table showing the photoionization and recombination times for some important ions contributing to the absorptions. This is at the face of the cloud and calculated using equilibrium approximation at a low flux state.}
    \label{time_tab}
\end{table*}

% such as C VI, O VII, and O VIII is $\sim 3 \times 10^3$ s, 

% For this model, C VI, O VII, and O VIII ions come to the final low flux equilibrium ion fraction value relatively quickly compared to the other ions, such as Si XII, Si XIV, Fe XV, and Fe XVI

\subsection{Level Populations}
This study examines the response of a gas cloud to changes in the radiation field. It is important to note that this response is not instantaneous and simultaneous throughout the cloud. Instead, there is a time delay in the response of the gas at different locations within the cloud. This delay is primarily influenced by the time it takes for the radiation to propagate through the cloud and the absorption processes that occur.

\begin{figure*}[ht]
\centering
\includegraphics[width=0.7\textwidth]{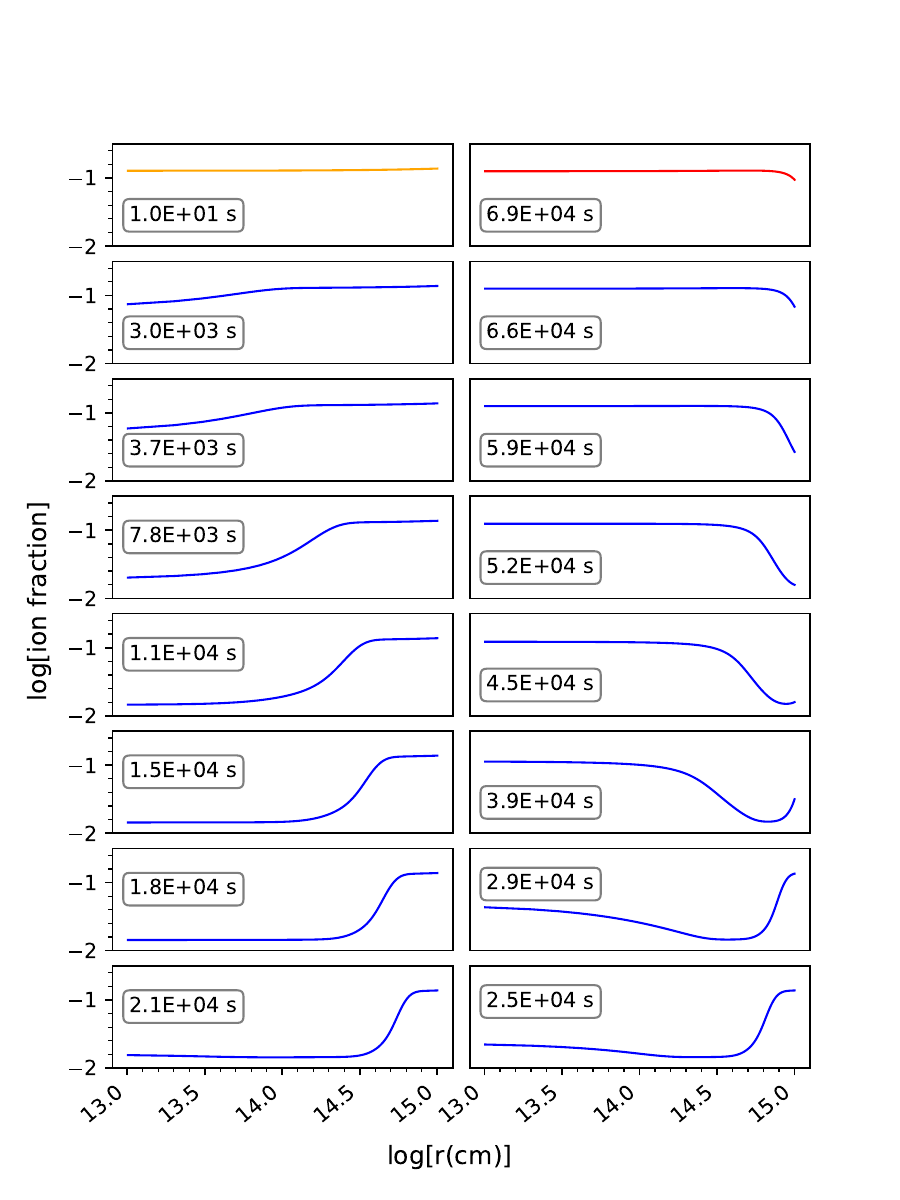}
\caption{Evolution of O VIII ion fraction profiles for model 2. The x-axis represents the depth r in the gas from the illuminated face, and the y-axis represents the ion fractions. The orange curve represents the initial lower flux equilibrium, while the red curve represents the final low flux equilibrium. Each blue curve corresponds to the quantity at an intermediate time during evolution.}

\label{ionfrac_vs_space}
\end{figure*}

To illustrate this, Figure \ref{ionfrac_vs_space} presents the ion fraction profile of oxygen (O VIII) at different times during the cloud's evolution. The ion fraction profile represents the relative abundance of O VIII ions within the cloud at various spatial locations. The value of the O VIII ion fraction is $\simeq 0.13$ in our simulation. This differs from the value at this same ionization parameter from \citet{kallman1982}, which was $\simeq 0.5$. The ion abundance curve for O VIII is shifted when these two calculations are compared.  The ion fraction vs. $\xi$ curves peaks at $\xi \simeq 15 $ erg cm s$^{-1}$ for the calculations shown here while it peaks at $\xi \simeq 50 $ erg cm s$^{-1}$ for the 1982 calculations.  For the standard XSTAR code plus database, the peak occurs at $\xi \simeq 25$ erg cm s$^{-1}$. Reasons for these differences include: differing energy grids,  the simplified version of the atomic database used in our current simulations, and other numerical factors.

At the initial time, when the radiation field starts to impact the cloud, the ion fraction profile of O VIII exhibits its initial distribution. As time progresses, the radiation field propagates through the cloud, reaching deeper into its interior. This leads to changes in the ionization state of the gas.

% Figure \ref{ionfrac_vs_space} shows the snapshots of the ion fraction profile in the cloud over a period of time during and after the flare. 
As depicted in Figure \ref{ionfrac_vs_space}, the ion fraction profile of O VIII at later times shows variations compared to the initial profile. The changes in the ion fraction occur gradually as the radiation field penetrates further into the cloud, resulting in a modified ionization state of oxygen. The time is shown in the box for each panel. The first left panel corresponds to the initial flux equilibrium ion fraction profile in the gas. This is before the gas gets flares. In the successive panels of the left column, the ion fractions are shown for increasing time. When the gas receives the flare, the ion fraction starts to change at the face of the cloud first. The ion fraction at the face of the cloud started to change at $\sim 3 \times 10^3$ s, gradually went down as flare went up, and reached a minimum at $\sim 2.1 \times 10^4$ s (the last panel of the left column) and start to increase as flare decreases and reach to the initial value again at time $\sim 3.9 \times 10^4$ s (sixth panel of the right column).

% The flare took $\sim 3.3 \times 10^4$ s at the back of the cloud. 
Propagation of the flare caused the ion fraction to decrease, reach the minimum, remain at a minimum for some time, and increase until it acquired the final low flux equilibrium value at all points in the cloud. The equilibration time at the back of the cloud is the sum of light travel ($\sim 3.3 \times 10^4$ s), flare ($\sim 3.0 \times 10^4$ s), and response time ($\sim 6.0 \times 10^3$ s), which comes to be  $\sim 6.9 \times 10^4$ s (the first panel of the right column). Therefore, it took $\sim 6.9 \times 10^4$ s for the whole gas to return to equilibrium. We note that the times in the panels are increasing on going down in the first column and increasing going up in the second column to commensurate with the shape of the flare.

\begin{figure*}
    \centering

    \includegraphics[width=0.8\textwidth]{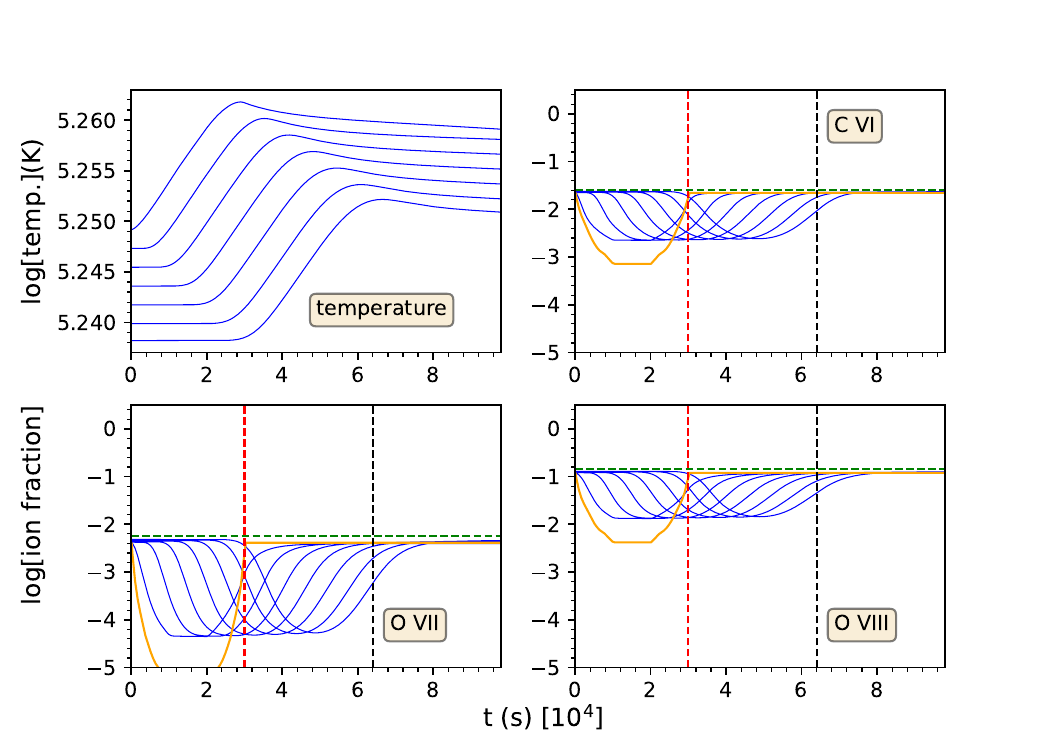}
    \includegraphics[width=0.8\textwidth]{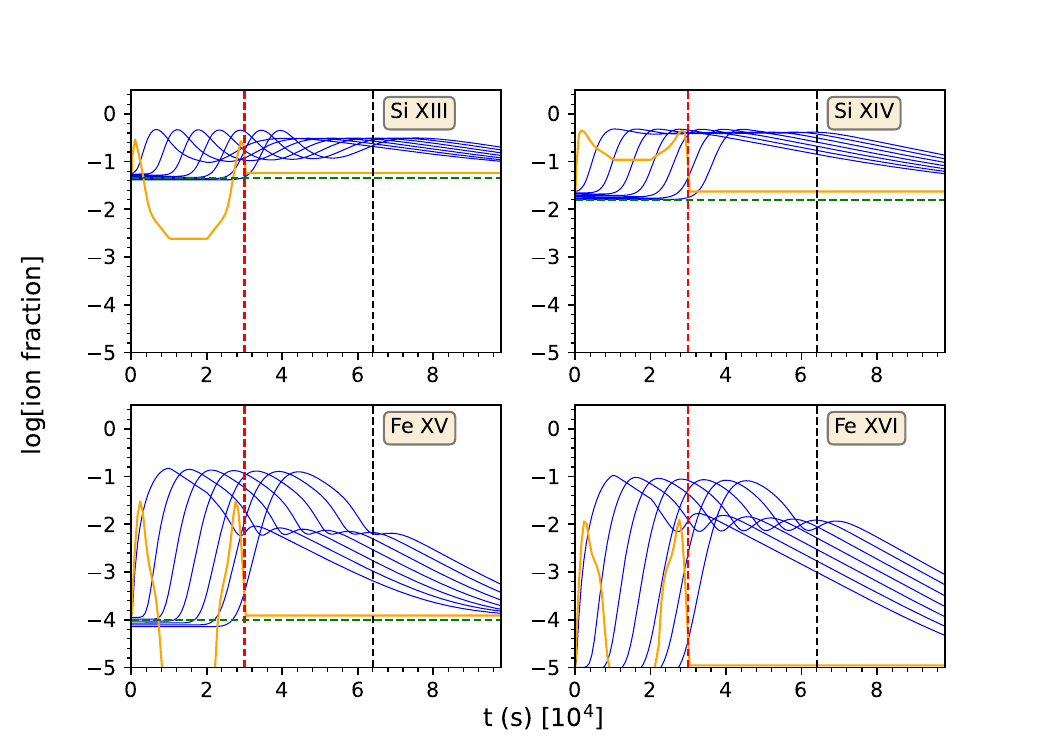}
    
    \caption{Time evolution of temperature (top left panel) and ion fractions (remaining all panels) for model 2. The x-axes are times in the unit of $10^4$ s. Each curve in all panels corresponds to a particular spatial point in the cloud. The one that evolves early is at the illuminated face of the cloud, and the latest is at the back of the cloud. The horizontal green line represents the equilibrium ion fraction for a low flux state. The red vertical lines represent the time of flare at the face. The blue vertical line, on the other hand, represents the time that it takes to propagate through the cloud.}
    \label{ionfr_vs_time}
\end{figure*}

The time evolution of the different ions at different spatial points in the cloud can be seen in figure \ref{ionfr_vs_time}. Each panel corresponds to a different ion except the first (top left) panel. This top left panel shows the temperature evolution in the cloud at different locations. The time dependent calculation predicts a temperature that evolved very little during the simulation. The temperature changed only by $\sim 10^4$ K for a flare, which increased in flux by a factor of ten. This temperature change is small compared to the change found if the gas were in equilibrium at the corresponding flux. The temperature changes by an order of magnitude in equilibrium approximation because the ionizing luminosity also changes by the same amount. The small increase in temperature in the time dependent calculation is attributed to the longer thermal time scale of low-density plasma. The ionizing source stays at a high luminosity state for $10^4$ s. This is much shorter than when it takes plasma to be thermalized. The thermal timescale of the plasma is about $\sim 10^5$ s for this model \citep{sadaula23}. In the rest of the panels, we have shown the evolution of representative ions such as C VI, O VII, O VIII, Si XII, Si XIV, Fe XV, and Fe XVI. The blue curves come from the time dependent calculation, whereas the orange lines come from the equilibrium approximation. The line that evolves earlier corresponds to the ion fraction at the illuminated face of the cloud, and the one that evolves slowest is at the back of the cloud; the rest are inside the cloud. In this plot, we have shown ion fractions at seven different points in the cloud, including the illuminated face and back of the cloud. The inner spatial points are at $\sim 1.7\times10^{14}$ cm, $\sim 3.4\times 10^{14}$ cm, $\sim 5.0\times 10^{14}$ cm, $\sim 6.8\times10^{14}$ cm, $\sim 8.5\times10^{14}$ cm from the illuminated face. We want to note that the physical size of the cloud for this model is $10^{15}$ cm. 
% This already shows that the evolution of the ions is not simultaneous throughout the cloud.

For this model, C VI, O VII, and O VIII ions come to the final low flux equilibrium ion fraction value relatively quickly compared to the other ions, such as Si XII, Si XIV, Fe XV, and Fe XVI. The time it takes for an ion to come to equilibrium corresponds to the time it takes the absorption spectrum to come to equilibrium at the relevant energies. More about the spectral evolution will be discussed in section \ref{waspec}.

The red dashed vertical line in all the panels indicates the time of the flare, which is $3\times 10^4$ s as seen in the light curve. The black dashed vertical line represents the time it takes for the flare to emerge completely from the cloud. This time is the sum of the flare time ($3\times 10^4$ s) plus the light travel time ($\sim 3.3\times 10^4$ s) of the cloud. The green dashed horizontal line is the low flux equilibrium value. All the panels show that, even if the flare has already passed, the ion fraction is still changing. However, the time scale to come to equilibrium is different for different ions. Ions such as C VI, O VII, O VIII, and Fe XV come to the equilibrium value by the time the simulation ends. However, ions such as Si XIII, Si XIV, and Fe XVI do not come all the way to the equilibrium values by the end of the simulation. They are close to the equilibrium when the simulation is ended; we estimate it would take an additional $\sim 5\times 10^4$ s to reach equilibrium for all these ions.
 
It is important to note that the equilibrium ion fraction values are different from those from the time dependent calculation. 
%This is completely different in some ions with intermediate ionization. 
Ions such as C VI, O VII, and O VIII have the same equilibrium ion fraction pattern as they do for time dependent models. They differ only in magnitude by some fraction. However, for ions such as Si XIII, Si XIV, Fe XV, and Fe XVI, the equilibrium ion fraction is qualitatively different from those from a time dependent calculation. The equilibrium ion fraction curves have an  `M shape' structure due to the flare. That is, at low flux early in the flare and late in the flare, the ion fractions increase with increasing flux and decrease with decreasing flux.   In the middle of the flare, the ion fractions decrease with increasing flux and increase with decreasing flux as the parent element is ionized to higher ion stages.  This behavior makes the time dependent modeling qualitatively different from equilibrium calculations.

The most significant feature of the behavior of some ion fractions vs. time is asymmetry. There is a tail-like structure in ions such as Si XIII, Si XIV, Fe XV, and Fe XVI. This implies when the flux is in the increasing phase of the flare, the ion fraction values change more rapidly than when the flare is in the declining phase.  This is important to keep in mind that time dependent ion fraction value at any time during evolution does depend on the condition at the previous times together with input parameters such as ionization parameter, time of the flare, density, etc. Once the gas experiences the flare and if the variability time of the flare is comparable to the photoionization and/or recombination timescale, a delay in the response occurs. This is just because the gas did not have enough time to respond to the change in the radiation field. The reason behind this is the behavior of the dominant microscopic time scale behind the increasing and declining phase of the flare. During the increasing phase, the gas starts to change its ionization state to a higher value, which is determined by photoionization, whereas recombination time plays a dominant role during the declining phase of the flare.  Furthermore, the recombination is dominated by dielectronic recombination for these ions, and this process is very temperature-sensitive.  As shown in paper 1, the temperature is determined by the thermal timescale, i.e., the cooling time, and for this model, the cooling time is longer than the recombination time. Thus, the gas remains stuck at a higher temperature late into the simulation, and at this temperature, the dielectronic recombination rate is less than it was at the initial equilibrium.  This results in the asymmetric ion fraction vs. time distribution for these ions. Successive ionization and recombination of the ions with the delayed response results in the complicated shape of ion evolution for these ions.

\subsection{Absorption Spectra}\label{waspec}

\begin{figure*}[ht]
\centering

\includegraphics[width=0.85\textwidth]{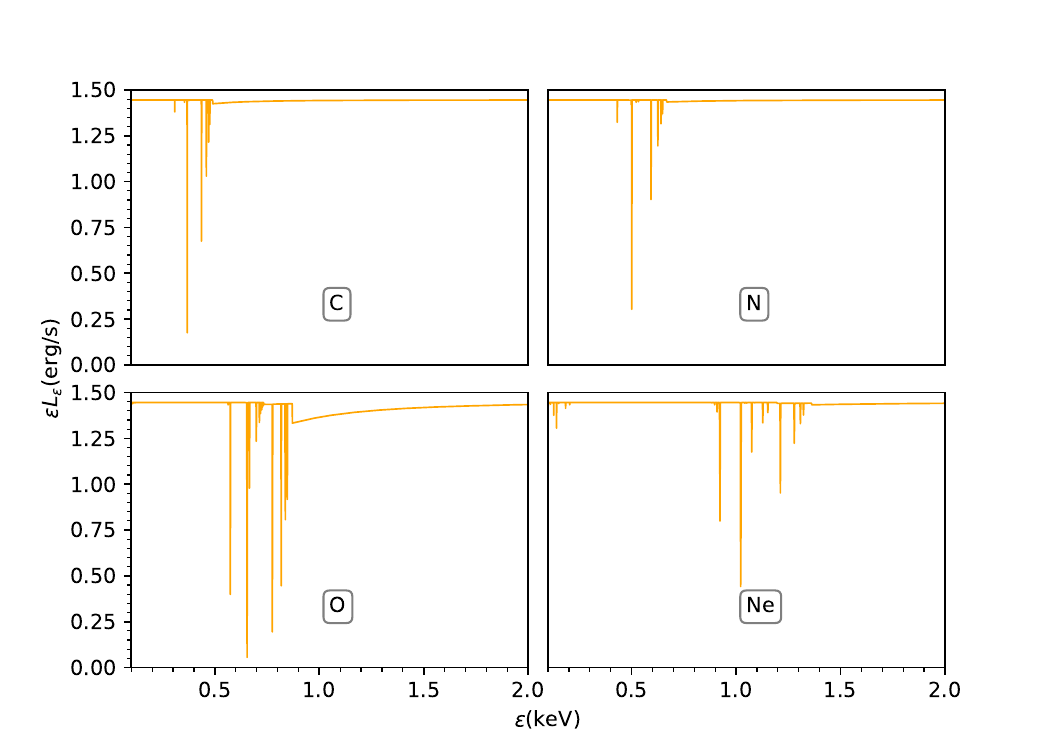}
\includegraphics[width=0.85\textwidth]{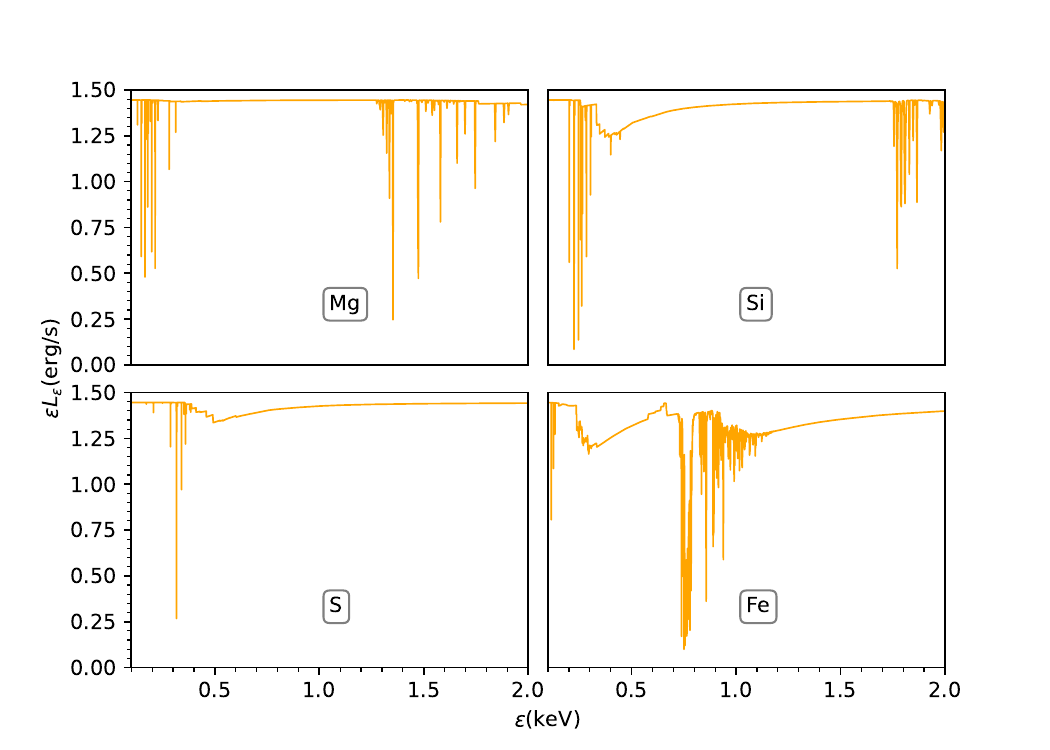}

\caption{Figure showing the elemental contribution in the absorption spectrum. The y-axes represent the transmitted luminosities ($\varepsilon L_\varepsilon$), and the x-axes represent the photon energies ($\varepsilon$). The units of $\varepsilon L_\varepsilon$ are $10^{43}$ erg s$^{-1}$. Each panel contains the low flux equilibrium spectrum caused by different elements. It is seen that O, Si, S, and Fe contribute most to the absorption spectra.}
\label{spec_elem}
\end{figure*}

Figure \ref{spec_elem} showcases the individual elemental contributions to the absorption spectrum. The model includes several elements such as H, He, C, N, O, Ne, Mg, Si, S, and Fe to simulate the response of a cloud to a flaring incident light curve. The figure highlights the relative importance of each element in contributing to the overall absorption spectrum. Among the elements considered, O (oxygen), Si (silicon), S (sulfur), and Fe (iron) dominate the absorption features in the spectrum. While the focus is on these key elements, it is important to note that the full spectrum encompasses absorption lines from all the included elements. The absorption features arise from a combination of edges and lines associated with the various ions. The energy range shown in the figure is up to 2 keV, as most ions primarily absorb radiation in the soft X-ray band. This range allows for a comprehensive representation of the absorption features resulting from the elemental contributions. The final model spectrum has three deep continuum absorption features at $\sim 0.3 $ keV, $\sim 0.75 $ keV, and $\sim 1 $ keV. These are denoted by first, second, and third absorption features in the spectrum described in the following section. In subsequent sections, the figure will be further explored to elucidate the specific absorption features, including edges and lines, present in the spectrum. This analysis will provide valuable insights into the ionization states, physical properties, and dynamics of the gas cloud under investigation.

\begin{figure*}[ht]
\centering

\includegraphics[height=22cm, width=16cm]{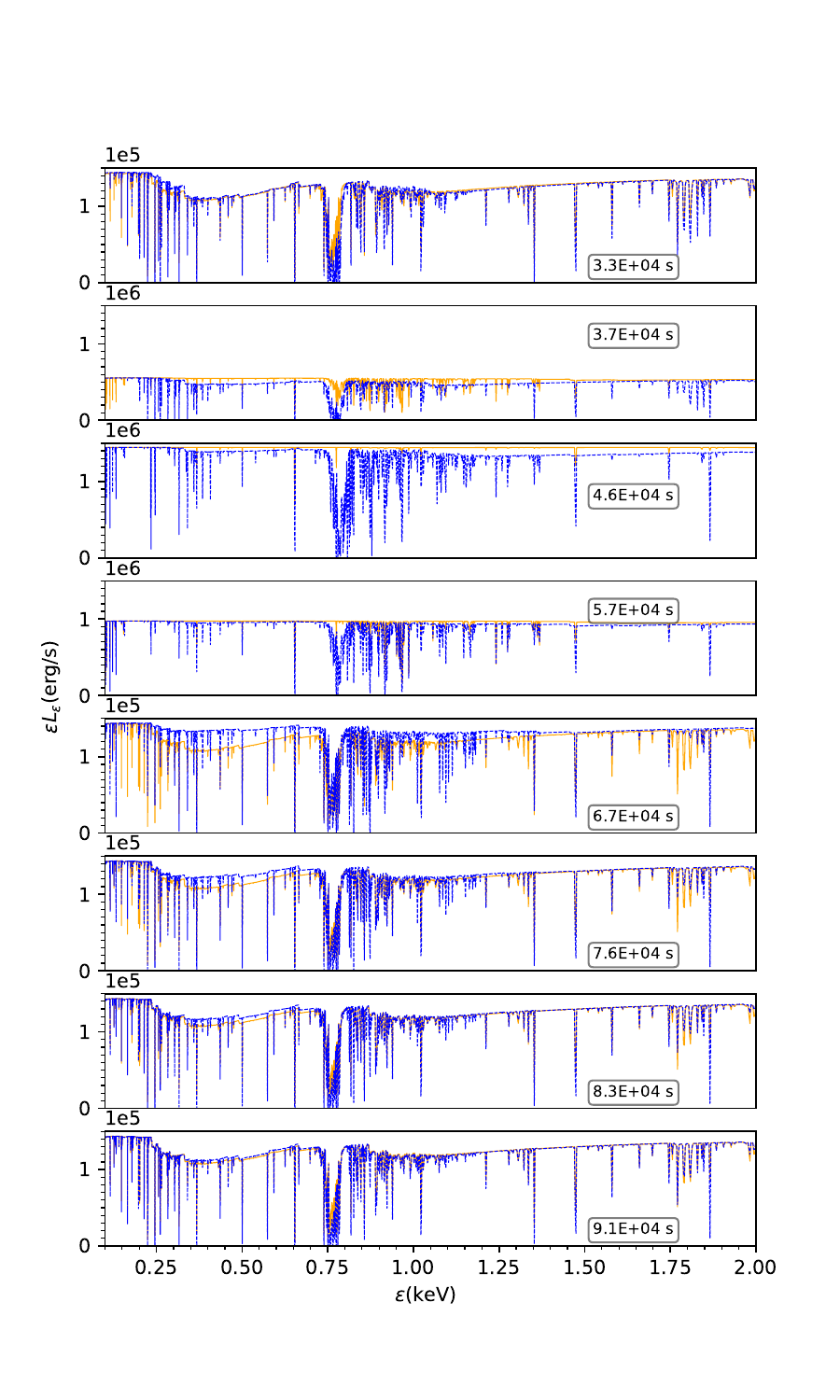}

\caption{Figure showing the evolution of the absorption spectra (blue curves) compared against the instantaneous flux equilibrium spectra (orange curves) for model 2. The y-axes represent the transmitted specific luminosities ($\varepsilon L_\varepsilon$), and the x-axes represent the photon energies ($\varepsilon$). The units of $\varepsilon L_\varepsilon$ are $10^{38}$ erg s$^{-1}$. Each panel corresponds to time as indicated by green vertical dashed and dotted lines in the light curve in figure \ref{lc_flare}. Note that the range of the y-axis is not the same for all panels. This is because the ionizing continuum of the source for this model changes by a factor of ten. The multiplicative factors for y-axes are shown at the top left of each panel.}  
\label{spectra_flare}
\end{figure*}

Figure \ref{spectra_flare} illustrates the temporal evolution of a modeled observed spectrum in response to a flare-like variation in the ionizing source. The figure consists of eight panels, each representing a specific time interval and displaying the corresponding absorption spectrum. The panels are divided into two sets of four. The first four panels depict the absorption spectrum during the passage of the flare through the cloud, while the last four panels represent the absorption spectrum after the flare has passed. The orange curve in each panel corresponds to an equilibrium spectrum generated using the instantaneous flux. This spectrum represents the expected absorption features without any temporal variations. The blue curves in the figure represent time dependent photoionization spectra at different points in time during the evolution. These spectra showcase the changes in the absorption features as the flux varies over time at the face of the cloud. By observing the figure, one can see how the absorption spectrum evolves over time in response to the changing flux from the ionizing source. The differences between the yellow and blue curves highlight the impact of temporal variations on the absorption features of the spectrum. 

Figure \ref{spectra_flare} comprises multiple panels, each representing a specific time corresponding to different vertical lines in the associated light curve. In the first panel, the time corresponds to the period before the flare reaches the back of the cloud. At this point, both the equilibrium and time dependent spectrum overlap significantly. The absence of significant divergence indicates that the spectra are similar, as the increased flux from the flare has not yet reached the cloud's back. This time interval aligns with the first vertical line in the light curve.

The second panel corresponds to a time of $\sim 3.7\times 10^4$ s (the sum of the light crossing time plus the time it took for the flare to reach this luminosity level), the luminosity of the source reaches $ \sim 4 \times 10^{44}$ erg s$^{-1}$. In this state, the equilibrium and time dependent spectrum begin to diverge. The equilibrium spectrum exhibits relatively less absorption compared to the time dependent spectrum. The first and third absorption features, observed in the time dependent spectrum, are nearly absent in the equilibrium spectrum. This difference arises from the higher ionization level of the gas due to the increased flux value.

The third panel represents the high luminosity state with source luminosity $ \sim 10^{45}$ erg s$^{-1}$, crucially showcasing the absence of absorption features in the equilibrium spectrum. In this state, the equilibrium spectrum shows minimal absorption, with most features being eliminated, except for a few weak absorption lines. In contrast, the time dependent spectrum continues to exhibit absorption features. The time corresponding to this particular state is $\sim 4.6\times 10^4$ s.

Finally, the fourth panel represents the point at which the source's luminosity begins to decrease, which corresponds to the fourth vertical dashed line from the left in the light curve \ref{lc_flare}.

The remaining four panels in Figure \ref{spectra_flare} depict the time evolution of the emergent spectra after the flare has completely emerged from the cloud. These spectra are compared to the instantaneous equilibrium spectra. In these panels, even after the flare has passed, the time dependent spectra continue to evolve, gradually approaching a low luminosity equilibrium absorption spectrum. The ongoing evolution is particularly noticeable in the edges and some lines in the spectra. The absorption lines below 0.25 keV, around 1.3 keV, and approximately 1.8 keV exhibit clear changes and evolution over time. The first absorption edge shows an apparent evolution pattern. This is because of the effect of longer equilibrium timescale of the ions such as Si XIII, Si XIV, Fe XV, and Fe XVI as shown in the figure \ref{ionfr_vs_time}. As the gas within the cloud settles into a low luminosity ionization equilibrium state, these lines continue to shift and evolve. The evolving nature of the spectra is indicative of the ongoing processes and adjustments occurring within the gas cloud. It highlights the time it takes for the ionization and excitation states of the gas to reach a stable equilibrium configuration after the passage of the flare.

According to the model, the cloud takes $\sim 9.1\times 10^4$ s to reach this low luminosity ionization equilibrium state. During this period, the time dependent spectra gradually converge toward the equilibrium absorption spectrum. By studying the time evolution of the emergent spectra in these panels, one can gain insights into the relaxation and settling processes of the gas cloud following the flare event. The changes in the absorption features, particularly in the edges and lines, provide valuable information about the ionization and the timescales involved in reaching a stable equilibrium state.

% This time dependent photoionization calculation can predict the ion fraction value at a given time and location in the cloud, as presented in Figure \ref{ionfr_vs_time}. 
The absorption features change rapidly during the evolution. This is primarily due to the ionization of Fe from M-shell ions such as Fe XV, Fe XVI, and Fe XVII, whose simple atomic structure does not have the rich inner shell spectra of the M-shell ions but contributes most to the absorption. It is also worth pointing out that the effects of saturation can be to make absorption features vary less with time than the abundances of their ions.   Thus, strong lines appear to be less variable than the ions.  Turbulent broadening plays a role in determining the importance of saturation.  In AGN outflows, typical turbulent velocities $\sim 10^3$ km s$^{-1}$, which will reduce the optical depths for many resonance lines. These effects are included in our simulation.

% In principle, the absorption lines are broadened mostly because of thermal broadening, doppler broadening caused by turbulence, and natural broadening

% \newpage
\section{Results: Model 3, flare model with single zone approximation}

The alternative approach of modeling warm absorber spectra using a single-zone approximation offers a computationally cheaper alternative to including radiative transfer in the calculations. In this method, the entire cloud is treated as a single zone, neglecting spatial variations and radiative transfer effects.   

% The most recent works that deploy this technique are carried out by \citep{rog22,luminari22}

\begin{figure*}[ht]
\centering

\includegraphics[height=22cm, width=16cm]{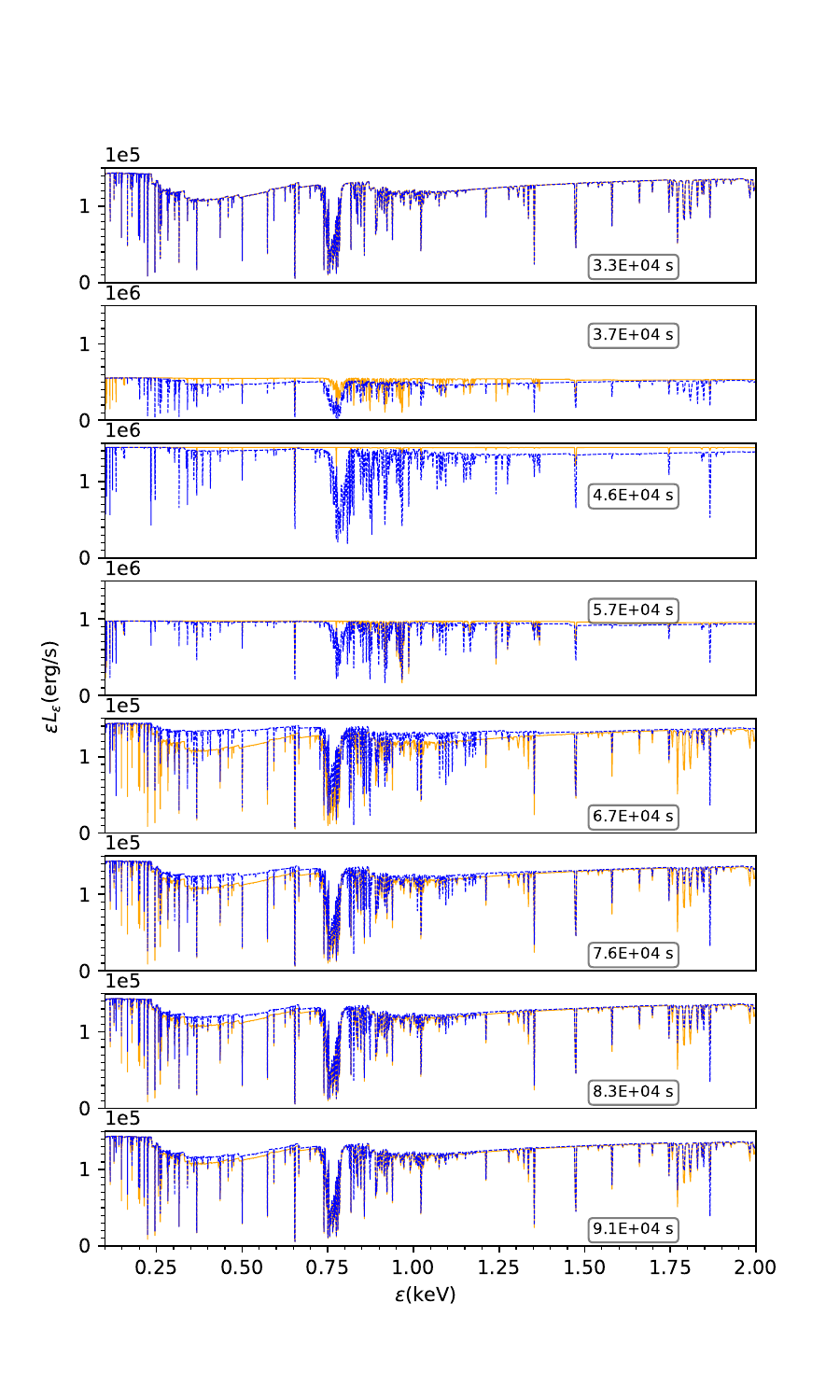}

\caption{Figure showing the evolution of the absorption spectra (blue curves) compared against the instantaneous flux equilibrium spectra (orange curves) for model 3. The y-axes represent the transmitted luminosities ($\varepsilon L_\varepsilon$), and the x-axes represent the photon energies ($\varepsilon$). The units of $\varepsilon L_\varepsilon$ are $10^{38}$ erg s$^{-1}$. Each panel corresponds to the time indicated by green vertical dashed and dotted lines in the light curve in figure \ref{lc_flare}. Note that the range of the y-axis is not the same for all panels. This is because the ionizing continuum of the source for this model changes by a factor of ten. The multiplicative factor for y-axes is written at the left top of each panel.}
\label{spec_singlezone}
\end{figure*}

The modeling begins by calculating the time dependent level population at the face of the cloud, solving the ion fraction, and heating and cooling equations for variable incoming radiation. It is then assumed that the evolution of the ion fraction pattern is the same throughout the cloud. This simplified approach produces spectra, as shown in Figure \ref{spec_singlezone}, which we refer to as model 3.

However, it is important to note that the timescale of evolution differs in the single-zone approximation compared to the multi-zone approach. By neglecting radiative transfer effects, important phenomena such as the absorption of ionizing radiation and geometric dilution are excluded. Consequently, the evolution of the ion fractions inside the cloud differs from that at the face of the cloud. Nevertheless, for thin gas clouds, the single zone approximation may provide sufficiently accurate results.   In this section, we examine this possibility.

Examining Figure \ref{spec_singlezone}, the top panel represents the initial state when the ionizing flux is at a low level, corresponding to the case before the light reaches the back of the cloud. As the increased flux passes through the entire cloud, the time dependent spectra start to diverge from the equilibrium spectra associated with the instantaneous flux equilibrium spectra. This divergence is observed in the second, third, fourth, fifth, and sixth panels. Eventually, the time dependent spectrum nearly overlaps with the equilibrium spectrum in the last panel, indicating that the ion fraction has reached the equilibrium value by that time.

In the single-zone approximation, the absence of radiative transfer calculations leads to the exclusion of the time delay caused by absorption.
% light passing through the cloud via processes such as absorption, emission, and scattering. {\bf I don't understand this.
 % {\bf We need to talk about it}
 Consequently, this method predicts the simultaneous change in optical depth at the face of the cloud and inside the cloud. This will unlikely happen in reality. There is some lag in the evolution of the ion at different points in the cloud due to the fact that the radiation needs to bleach through the cloud. This effect is particularly evident in the absorption spectrum and becomes more pronounced with a greater physical thickness of the gas. In the present model, the spectrum does not converge well to the final low flux equilibrium spectrum, as observed in the multi-zone model. This is more visible at the first absorption feature. The absorption lines at $\sim 1.8 $ keV also did not quite overlap with the absorption lines that come from equilibrium approximation, which is given in orange color. This discrepancy arises from the different ion fractions and their evolution at the front and back of the absorber, as depicted in Figure \ref{ionfr_vs_time}. Specifically, ions such as Si XIII, Si XIV, Fe XV, and Fe XVI exhibit different ion fraction values at the front and back of the cloud during specific stages of evolution.

\begin{figure*}[ht]
\centering

\includegraphics[width=0.9\textwidth]{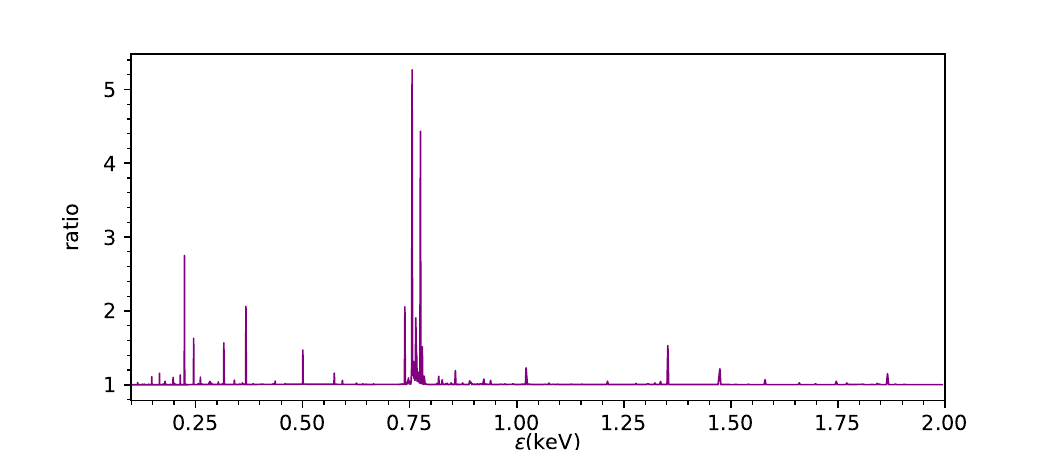}

\caption{Figure showing the ratio of emergent specific luminosity for model 3 ( single zone ) to model 2 (multiple zones) at a time $9.1 \times 10^4$ s, which are the last panels of the figures \ref{spectra_flare} and \ref{spec_singlezone}. This demonstrates the difference between the two modeling approaches, and model 3 under-predicts the strengths of absorption features systematically. }  
\label{comp}
\end{figure*}

The discrepancy between single-zone and multiple-zone approach modeling is illustrated by figure \ref{comp}. This figure shows the ratio of the emergent specific luminosity for single-zone to multiple-zone approximations. The ratio is expected to be one if both techniques produce the same results. However, the figure shows that the ratio is greater than one for some specific wavelengths where the absorption lines appear in the spectrum. This means the depths of the absorption lines in the single-zone model are underestimated compared with the multiple-zone model. In the case of the multiple zones, the cloud is divided into many spatial zones, and the absorption by each zone is calculated to find a radiation field throughout the cloud, which is used to find the ionization state. This process incorporates absorption and geometric dilution, resulting in different ion fractions from the single zone approximation deeper in the cloud. The effect is significant for those energy values where opacity is high, and this is why the two methods diverge more in the lines than the continuum. The spikes in figure \ref{comp} correspond to the ratio of absorption lines. The ratio is highest at 0.75 keV. This energy corresponds to the absorption lines from L-shell iron ions.

In summary, the single-zone approximation offers a computationally efficient approach for modeling warm absorber spectra. While it neglects radiative transfer effects and the associated time delays, it can still provide valuable insights, particularly for thin gas clouds. The differences observed in the ion fraction evolution and absorption spectra between the front and back of the cloud emphasize the limitations of this approach compared to more detailed multi-zone models.

\newpage
\section{Conclusion}
Understanding warm absorbers is essential for gaining insights into active galactic nuclei and their surrounding environments \citep{kallman2019,blustin2005,san2018,rey95}. When the illuminating flux undergoes variations comparable to the ionization, recombination, and propagation timescales in warm absorber clouds, relying solely on photoionization and thermal equilibrium assumptions can yield misleading interpretations when analyzing observations. This study addresses these challenges by incorporating AGN source variability and employing non-equilibrium photoionization calculations to model the outflow. The main focus of this research is on the high-resolution spectrum and its temporal evolution, a crucial aspect for forthcoming missions like XRISM and Athena. The developed model is particularly suited for analyzing the soft X-ray band (0.1 - 2 keV). By fitting the evolving spectrum to a variable ionizing continuum, valuable information about the warm absorber's properties, such as density, location, and ionization state, can be extracted and constrained. This paper introduces the concept of employing time dependent photoionization calculations to accurately track changes in the absorber's ionization state over time. For the specific light curve considered in this study (model 1), the equilibration time for the spectral response is estimated to be $\sim 5.8 \times 10^6$ seconds.

In this study, we have incorporated two fundamental types of variability, namely step-up and flare, which closely resemble real-world scenarios. Sudden increases in source luminosity due to accretion rise are common in AGN \citep{vag2016,utt2005,mchardt2004,ponti2012}, and we have extensively examined and presented the properties of the warm absorber during flare events. The microphysical timescales play a significant role in responding to the variation of the ionizing continuum and depend on factors such as density. At higher densities, the gas tends to remain closer to equilibrium, while at lower densities, the gas is less responsive to changes in illumination. For intermediate densities, simulations like the ones presented in this study are crucial. By carefully tracking the ionization state and absorption spectra, we can deduce vital information, particularly the density, about the warm absorbers. This knowledge contributes to our understanding of AGN morphology, dynamics, and the role of warm absorbers in AGN feedback. This research lays the foundation for further advancements in this field. For the flare model (model 2), we observed that it takes approximately $9.1 \times 10^4$ seconds for the spectrum to settle down to a state of low flux equilibrium.

Warm absorber spectra are commonly studied and analyzed by fitting multiple equilibrium components, with each component parameterized by a single ionization parameter \citep{reynolds1997,laha14,kas02,blustin2002,beh03}.  Past time dependent photoionization studies of warm absorbers have overlooked a crucial element: the effect of time dependent radiative transfer.  That is, the coupling between the time dependent ionization of the gas, and hence its opacity, with the transport of the radiation. In our work, we have taken coupled time dependent photoionization and radiative transfer into account by solving the time dependent transfer equation through multiple spatial zones of ionized gas. Although computationally demanding, this approach is more representative of the actual behavior within the warm absorber. 
%In our work, we address this issue by carefully dividing the absorber cloud into distinct zones, each characterized by a different ionization state. This is accomplished through the inclusion of time dependent radiative transfer equations in our computations. 
Our study demonstrates that the choice between single-zone and multiple-zone models has a significant impact on the spectral evolution of the specific model considered here. The disparity becomes even more pronounced with higher column density. A comparison of models with and without time dependent radiative transfer shows that they undergo distinct spectral evolutions, highlighting differences in timescales and patterns. By accounting for the propagation of the flare through the cloud and incorporating the associated propagation time, the multiple-zone model provides a more realistic representation of the temporal evolution and smoothing of the flare as it interacts with the gas cloud.

\section{acknowledgement}
We thank Dr. Alan Hindmarsh from Lawrence Livermore National Laboratory for his help and valuable suggestions regarding the ordinary differential equation solver package. We are grateful to the anonymous referee for their constructive feedback.
This work was partly funded via NASA grant 17-ATP17-0113 through the Astrophysics Theory Program.
% This work was supported by grants through the NASA Astrophysics Theory Program.

\appendix
\section{Theory}\label{sec2}

Our models consist of the solution of the coupled equations describing level populations, temperature, and radiation field. paper1 \cite{sadaula23} presents all these equations' details. Here, we briefly repeat some of them.

\subsection{Level Populations}

The population of an atomic level obeys the kinetic equation involving the atomic rates into and out of the level:  

\begin{equation}\label{levpop}
    \frac{dn_{i,X}}{dt}=\sum_{j=1}^p n_{j,X} R_{ji}-\sum_{k=1}^p n_{i,X} R_{ik}
\end{equation}
and the equation of number conservation
\begin{equation}\label{chargebalance}
    \sum_{i=1}^p n_{i,X} = x n_H   
\end{equation}
where $n_{i,X}$ are the level populations in $i^{th}$ energy level of element X in units  cm$^{-3}$, which is equal to the product of the fractional elemental abundance and total hydrogen number gas density, $n_H$. $ x$ is the fractional elemental abundance of species $X$ relative to hydrogen. $R_{ji}$ is the transition rate from the $j^{th}$ to $i^{th}$ energy level contributed by atomic processes, including photoexcitation, photoionization, collisional excitation, collisional ionization, recombination, charge transfer, and radiative decay. The value of rates for $R_{ij}$ are taken from XSTAR. $p$ is the total number of energy levels considered for the particular element. As discussed in section 3, we adopt 
a simplified level structure so that, for an element with atomic number Z, p=$2Z+1$. Here and in what follows, we adopt the values of the atomic constants (e.g., photoionization cross-sections, atomic energy levels, collision rate coefficients) from the {\sc xstar} database \citep{bau01}.

\subsection{Temperature}
The temperature variation in the gas is given by this differential equation,
\begin{equation}\label{tempequn}
\frac{dT}{dt}=\frac{2}{3kn_{t}}(\Gamma^{heat}-\Lambda^{cool}-\frac{3}{2}kT\frac{dn_{t}}{dt})
\end{equation}
where $T$ is the temperature, $\Gamma^{heat}$ and $\Lambda^{cool}$ are total heating and cooling rates in units of ${\rm erg}\ {\rm s}^{-1}\ {\rm cm}^{-3} $, $k$ is the Boltzmann constant, and $n_t$ is the total number density of particles, i.e., the sum of all electrons, ions, and neutral atoms. The last term on the right-hand side of this equation is the rate of change in the internal energy density associated with the rate of change of the total number of free particles in the gas. This is negligible when the gas is highly ionized; otherwise, it explicitly couples the level populations with the temperature equation.

\subsection{Time Dependent Radiative Transfer}
The equation describing the time evolution of the radiation field in spherical symmetry is \citep{hat76,gar13},
\begin{equation}\label{rad}
    \frac{1}{c}\frac{\partial{L_\varepsilon(R,t)}}{\partial t}+ \frac{\partial{L_\varepsilon(R,t)}}{\partial R} = 4 \pi R^2 j_\varepsilon-\kappa_\varepsilon(R,t) L_\varepsilon(R,t) \end{equation}
where $R$ is the distance in the cloud from the ionizing source, $L_\varepsilon(R,t)$ is the specific luminosity of the ionizing source in erg s$^{-1}$  erg$^{-1}$, $j_\varepsilon$ is the local emissivity in erg s$^{-1}$ cm$^{-3}$ erg$^{-1}$, and $\kappa_\varepsilon(R,t)$ is the total extinction coefficient in cm$^{-1}$. In our calculations in this paper, the extinction comes from only absorption and we do not include emission, so the equation becomes:

\begin{equation}\label{transfer}
    \frac{1}{c}\frac{\partial{L_\varepsilon(R,t)}}{\partial t}+ \frac{\partial{L_\varepsilon(R,t)}}{\partial R} = -\kappa_\varepsilon(R,t) L_\varepsilon(R,t)
\end{equation}

The rest of the equations for different time scales and their comparison are mentioned in detail in our previous work \citep{sadaula23}. We have generated a mini version of the full XSTAR database, including only three levels; one ground, one excited, and one ionized level. This newly generated database is then tested with multiple methods and verified to produce adequate accuracy in our time dependent computation. The readers are encouraged to review our previous paper \citep{sadaula23}. The recombination, photoionization, and light travel time play an important role in the evolution and equilibration of the gas when it experiences variable ionizing radiation. The recombination time is the most important of them, which essentially depends on the gas density.

\section{Numerical approach}\label{if0}
The set of ordinary differential equations for our problem governs level populations, temperature, electron number density, and radiation field.  In the general case, this system of equations can have terms that cover a large range of values, owing to the diverse physical processes they describe.  This system is, therefore,  'stiff' \citep{coo69}; that is, they correspond to a set of variables (solutions of differential equations)  changing from a much faster rate to a much slower rate.  Some processes are important for small timescales, while others dominate for long timescales. In order to solve these equations, we implement the well-known ordinary differential equation solver, DVODE \citep{bro89}.

\bibliography{tdp}{}
\bibliographystyle{aasjournal}
\end{document}